\RequirePackage{fix-cm}
\documentclass[onecolumn,epjc3,longbibliography]{svjour3}  

\smartqed  
\RequirePackage{graphicx}
\journalname{Eur. Phys. J. C}

\usepackage{geometry}
\usepackage[dvipsnames]{xcolor}
\usepackage{amssymb}
\usepackage{cite}
\usepackage{amsmath}
\usepackage[normalem]{ulem}
\usepackage{soul}
\usepackage{graphicx}
\usepackage{dcolumn}
\usepackage{bm}
\usepackage{epstopdf}
\usepackage{mathrsfs}
\usepackage{amssymb,amsmath,amsfonts,latexsym}
\usepackage{nicefrac}
\usepackage[colorlinks=true,linktocpage=true,linkcolor=blue,citecolor=blue]{hyperref}
\usepackage[utf8]{inputenc}
\usepackage{lipsum}
\usepackage{slashed}
\usepackage{mathtools}
\usepackage{multirow}
\usepackage{caption}
\usepackage{subcaption}
\usepackage{lineno}

\def\del{\partial}

\def\sst{\scriptscriptstyle}

\long\def\comment#1{ }

\def\0{{\boldsymbol 0}}

\def\r{{\boldsymbol r}}

\def\tform{{t_\text{f}}}
\def\tdecoh{t_\text{d}}

\def\Kc{{\cal K}}

\def\pt{p_{\sst T}}
\def\pT{p_{\sst T}}
\def\pTz{p_{\sst T,0}}

\newcommand{\beq}{\begin{eqnarray}}
\newcommand{\eeq}{\end{eqnarray}}

\newcommand{\be}{\begin{eqnarray*}}
\newcommand{\ee}{\end{eqnarray*}}

\newcommand{\rmd}{{\rm d}}
\newcommand{\dd}{{\rm d}}
\newcommand{\rme}{{\rm e}}
\newcommand{\rmi}{i}

\def\glu{{\rm g}}
\def\glug{{\rm gg}}
\def\gq{{\rm gq}}
\def\qg{{\rm qg}}
\def\qq{{\rm qq}}

\def\S{{\rm S}}

\newcommand{\nn}{\nonumber\\ }

\newcommand{\qhat}{\hat{q}}
\def\Raa{R_\mathrm{AA}}

\begin{document}

\title{Multi-partonic medium induced cascades in expanding media
}


\author{ Souvik Priyam Adhya\thanksref{e1,addr1,addr1a}
        \and
        Carlos A. Salgado \thanksref{e2,addr2}
        \and
        Martin Spousta\thanksref{e3,addr1} 
        \and
        Konrad Tywoniuk\thanksref{e4,addr3} 
}

\thankstext{e1}{souvik.adhya@ifj.edu.pl}
\thankstext{e2}{carlos.salgado@usc.es}
\thankstext{e3}{martin.spousta@mff.cuni.cz}
\thankstext{e4}{konrad.tywoniuk@uib.no}


\institute{Institute of Particle and Nuclear Physics, Faculty of Mathematics and Physics, Charles University, V Hole\v sovi\v ck\'ach 2, 180 00 Prague 8, Czech Republic\label{addr1}
           \and
           Institute of Nuclear Physics, Polish Academy of Sciences,\\
  ul.\ Radzikowskiego 152, 31-342 Krakow, Poland \label{addr1a}
           \and
           Instituto Galego de F\'isica de Altas Enerx\'ias IGFAE, Universidade de Santiago de Compostela, E-15782 Galicia-Spain \label{addr2}
           \and
          Department of Physics and Technology, University of Bergen, 5007 Bergen, Norway\label{addr3}
}

\date{Received: date / Accepted: date}

\maketitle

\begin{abstract}

  Going beyond the simplified gluonic cascades, we introduce both gluon and quark degrees of freedom for partonic cascades inside the medium. We then solve the set 
of coupled evolution equations numerically with splitting kernels calculated for static, exponential, and Bjorken expanding media to arrive at medium-modified parton 
spectra for quark and gluon initiated jets. Using these, we calculate the inclusive jet $\Raa$ where the phenomenologically driven combinations of quark and gluon 
jet fractions are included. Then, the rapidity dependence of the jet $\Raa$ is examined. We also study the path-length dependence of jet quenching for different 
types of expanding media by calculating the jet $v_2$. Additionally, we study the sensitivity of observables on effects from nuclear modification of parton 
distribution functions, vacuum-like emissions in the plasma, and the time of the onset of the quenching. All calculations are compared with recently measured data.

\end{abstract}

\section{Introduction}

One of the most striking phenomena that was predicted \cite{Bjorken:1982tu,Gyulassy:1990ye,Wang:1994fx} to occur in relativistic collisions of heavy nuclei and that was confirmed by numerous experiments at RHIC \cite{Adcox:2001jp,Adler:2002xw,Adler:2002tq} and 
the LHC \cite{CMS:2012aa,Aad:2012vca,Aad:2015wga,Aamodt:2010jd,Aad:2014bxa,Adam:2015ewa,Khachatryan:2016jfl} is the suppression of high-$p_T$ hadron and jet production, generally referred to as jet ``quenching'', for reviews see \cite{dEnterria:2009xfs,Majumder:2010qh,Mehtar-Tani:2013pia,Blaizot:2015lma}. This phenomenon is generally understood as a consequence of energy loss of partons traversing the hot and dense QCD medium consisting of the soft degrees of freedom released in the collisions. Over the last two decades, the theoretical description of jet quenching evolved from single-parton energy loss to include multi-parton coherence effects and a deeper understanding of thermalization effects, for reviews see \cite{CasalderreySolana:2007zz,Mehtar-Tani:2013pia,Blaizot:2015lma,Cao:2020wlm}, paralleled by a rapid development of Monte Carlo models \cite{Lokhtin:2005px,Zapp:2008gi,Armesto:2009fj,Renk:2010zx,Young:2011ug,Casalderrey-Solana:2014bpa,He:2015pra,Cao:2017zih,Putschke:2019yrg,Caucal:2019uvr,Ke:2020clc}.

Not all observables have the same sensitivity to the degree of complexity implemented in the theoretical description and therefore not the same sensitivity to the 
underlying physics which is modeled. Consequently, it is important to study the impact of individual aspects of complex modeling on individual observables. In this 
paper, we focus mainly on the canonical jet spectrum in lead-lead collisions at the LHC and the related nuclear modification factor, $\Raa$.
These observables are mainly sensitive to the energy lost by the leading particles in the jet, due to the bias from the steeply 
falling hard cross-section of jet production \cite{Baier:2001yt}. To quantify in better detail the dependence on path length of jets, we also focus on the azimuthal 
asymmetry of jet production quantified in the second harmonic coefficient of the jet multiplicity distribution, $v_2$.
  All these observables have been analyzed in various frameworks before. However, in this paper we present a novel implementation of the effects of expanding medium which 
allows to quantify the sensitivity to the plasma evolution (for related work, see also \cite{Caucal:2020uic}), 
and we present the calculation of the inclusive jet suppression evaluated differentially in the full set of basic kinematic quantities: transverse momentum, rapidity, 
and the azimuth.

In a recent work \cite{Adhya:2019qse}, we have presented a study of the distribution of medium-induced gluons and the jet calculated using the evaluation of 
in-medium evolution with splitting kernels derived from the gluon emission spectra for expanding profiles. Scaling behavior of splitting kernels was derived for 
low-$x$ and high-$x$ regimes in the asymptote of large times and its impact on the resulting jet $\Raa$ was discussed.\footnote{Since, at that point, we were 
including only gluons, we were referring to the phenomenological jet suppression factor as $Q_{AA}$.} For the full phase space of the radiation, the scaling of jet 
$\Raa$ with an effective quenching parameter was introduced.

In this paper, the main aim is to study the impact of the medium expansion on observed jet suppression within more realistic description of the jet production. 
The description is improved by:
  1) by including both gluon and quark degrees of freedom for partonic cascades into the evolution equations; 
  2) by improving on the description of the initial parton spectra.
  This should provide more realistic calculation of nuclear modification factor and more realistic values of quenching parameter, $\hat{q}$, to be extracted from the data. 
  Within this description we evaluate the rapidity dependence of $\Raa$ and path-length dependence of the energy loss (via jet $v_2$) and study if measurements of 
inclusive jet suppression differential in rapidity and azimuthal angle are sensitive to the way how the medium expands. 
  Furthermore, we study the sensitivity of observables on effects from nuclear modification of parton distribution functions, effects from vacuum-like emissions in 
the plasma, and sensitivity on the time of the onset of the quenching.

The paper is organized as follows. In Sec.~\ref{sec:spectra} we summarize the single parton emission spectra and the in-medium splitting rates for possible flavor 
combinations at the leading order. Next, we introduce the coupled evolution equations to arrive at the medium evolved parton spectra with the in-medium splitting 
functions as the kernels of the evolution. In Sec.~\ref{sec:jetsup}, we first calculate jet $\Raa$ from moments of medium evolved partonic spectra for quark and gluon initiated jets. Then, in the next four subsection we analyze the transverse momentum dependence, initial-time dependence, rapidity dependence, and path-length dependence of the jet suppression.
We summarize and conclude in Sec.~\ref{sec:summary}.
\section{In-medium emissions for different media profiles}
\label{sec:spectra}

\subsection{Coupled parton cascade evolution equations}
Medium-induced radiation was traditionally considered in the context of soft gluon emissions 
\cite{Baier:1996kr,Baier:1996sk,Zakharov:1996fv,Zakharov:1997uu}. Naturally, the turbulent nature of the cascade was first discussed in the context of gluon fragmentation \cite{Blaizot:2013hx,Blaizot:2015jea}. Nevertheless, because of the important role played by quasi-democratic splittings, it is natural to consider the interplay of quark and gluon degrees of freedom in the cascade \cite{Mehtar-Tani:2018zba,Schlichting:2020lef}. From our point of view, this is also a natural demand when considering quenching effects for a realistic initial hard spectrum of initial partons, which is dominated by gluons at low- and intermediate-$p_T$ and by quarks at high-$p_T$. For consistency, one should also consider the feedback from gluon to quark-antiquark splittings.

The main quantity of interest is the single-inclusive in-medium parton distributions, defined as
\beq
D_\rmi(x,\tau) \equiv x \frac{\rmd N_\rmi}{\rmd x},
\eeq
where $x=\omega/E$ denotes the energy fraction carried by a parton of frequency $\omega$ with respect to the initial energy $E$ of the original parton\footnote{Here, we concretely refer to the light-cone momentum fraction $E \equiv p^+ = (p^0 + p^3)/2$ and $p^- = p^0-p^3$.} and $i$ refers to the parton flavor. This is measured at the dimensionless ``time'' $\tau \equiv \sqrt{\hat q_0/E}L$, where  $L$ is the in-medium path-length.\footnote{Sometimes the evolution variable is defined with an additional $\alpha_s$, so that $\tau= L/t_{\rm stop}$ , where $t_{\rm stop} = \bar \alpha^{-1} \sqrt{E/\hat q}$ is the stopping time of a jet with energy $E$. Because we are including finite-size effects, we have instead chosen to keep the coupling constant explicitly in the splitting functions.} Here, $\hat q_0$ is the value of  the jet quenching coefficient $\hat q$ at an initial reference time. Hence, the distribution $D(x,\tau)$ also intrinsically depends on this time scale. For a medium with constant or exponentially decaying density the natural initial time is simply $t=0$ and this dependence is trivial. However, this is not the case for power-like decaying medium profiles, which are associated with a finite starting time. This will be discussed in more detail below.
 
Concretely, we  focus on the gluon $D_g(x)$ and quark singlet distributions $D_S(x) \equiv \sum_f \big[D_{q_f}(x) + D_{\bar q_f}(x) \big]$, where the sum runs over all active flavors, which describes the energy distribution of quarks inside a jet.
The evolution equations for the inclusive in-medium parton distribution can be written as \cite{Mehtar-Tani:2018zba}
\begin{align}
\label{eq:evol-eq-glu}
\frac{\del }{\del \tau } D_\glu\left(x,\tau\right)&=\int_{0}^{1} dz \, \Kc_\glug(z) \left[ \sqrt{\frac{z}{x}} D_\glu\left(\frac{x}{z},\tau \right) \Theta(z-x) - \frac{z}{\sqrt{x}} D_\glu(x,\tau) \right] \nn
&-   \int_{0}^{1} \rmd z \,\Kc_\qg(z)  \frac{z}{\sqrt{x}}\, D_\glu\left(x ,\tau \right) +\int_0^1 \rmd z \, \Kc_\gq(z)  \,  \sqrt{\frac{z}{x}} \,D_\S \left(\frac{x}{z},\tau \right) \,,\\
\label{eq:evol-eq-S}
\frac{\del }{\del \tau } D_\S\left(x,\tau\right)&= \int_{0}^{1} dz \, \Kc_\qq(z) \left[ \sqrt{\frac{z}{x}} D_\S\left(\frac{x}{z}, \tau \right) \Theta(z-x) - \frac{1}{\sqrt{x}} D_\S(x,\tau) \right] \nn
& + \int_{0}^{1} dz \, \Kc_\qg(z) \, \sqrt{\frac{z}{x}} D_\glu\left(\frac{x}{z} ,\tau \right) \,.
\end{align}
Each equation involves a positive gain term and a negative loss term which describe the production of a parton with energy fraction $x$ from the splitting of a parent parton with energy fraction $x/z$ and the decay of a parton with energy fraction $x$ into softer fragments with fractions $zx$ and $(1-z)x$, respectively.

The splitting kernels $\Kc_{ij}$ quantify the rate of splittings per unit evolution time $\tau$. They can be expressed as
\beq
\Kc_{ij} (z,\tau) \equiv \frac{\rmd I_{ij}}{\rmd z \, \rmd \tau} \,,
\eeq
where $\rmd I_{ij}/\rmd z$ is the medium-induced emission of parton $i$ from parton $j$, carrying a longitudinal momentum fraction $z$.
  The main purpose of this work is to explore the evolution of the cascades for expanding media. Below, we focus on three main scenarios:
\begin{description}
\item[Static medium:] In a static medium, $\hat q(t) = \hat q_0 \Theta(L-t)$, where $L$ is the medium length. The splitting rate reads
\beq
\label{eq:rate-static}
\mathcal{K}^{\rm stat}_{ij}(z,\tau)  =  \frac{\alpha_s}{2\pi} P_{ij}(z) \kappa_{ij} (z)\, \text{Re} \left[(i-1) \tan \left(\frac{1-i}{2} \kappa_{ij} (z) \tau \right) \right]\,.
\end{eqnarray}
For the case of soft gluon emissions, i.e. $1-z,z \ll 1$, in a static medium, the splitting function is given as,
\beq
\label{eq:rate-static}
\mathcal{K}^{\rm stat (soft)}_{ij}(z,\tau)  \simeq  \frac{\alpha_s}{2\pi} P_{ij}(z) \kappa_{ij} (z)\,.
\end{eqnarray}

\item[Exponentially decaying medium:] For exponentially decaying media the profile of the jet quenching parameter is given by $\hat q(t) = \hat q_0 {\rm e}^{-t/L}$. The splitting rate reads,
\beq
\label{eq:rate-expo}
\mathcal{K}^{\rm exp}_{ij}(z,\tau)  = \frac{\alpha_s}{\pi} P_{ij}(z) \kappa_{ij}(z) \,\text{Re} \left[ (i-1) \frac{J_1\big((1-i)\kappa_{ij}(z) \tau \big)}{J_0\big((1-i)\kappa_{ij}(z) \tau \big)} \right] \,.
\eeq

\item[Power-law decaying medium:] For power-law decaying media, the jet quenching parameter is defined as
\beq
\hat q(t) = \begin{cases} 0 & {\rm for } \quad t<t_0 \,, \\ \hat q_0 (t_0/t)^\alpha & {\rm for} \quad t_0 < t < L+t_0 \,, \\ 0 & {\rm for} \quad L+t_0 < t \,,\end{cases}
\eeq
with $\alpha =1$ corresponding to the Bjorken expansion \cite{Baier:1998yf}. The splitting rate reads in this case,
\begin{align}
\label{eq:rate-bjorken}
\mathcal{K}^{\rm BJ} _{ij}(z,\tau,\tau_0) &= \frac{\alpha_s}{2\pi} P_{ij} (z) \kappa_{ij} (z)\sqrt{\frac{\tau_0}{\tau + \tau_0}} \nn
&\times \text{Re} \left[ (1-i) \frac{J_1(z_L) Y_1(z_0) - J_1(z_0) Y_1(z_L) }{J_1(z_0) Y_0(z_L) - J_0(z_L) Y_1(z_0)} \right],
\end{align}
where
\begin{align}
z_0 &= (1-i) \kappa_{ij} (z) \tau_0 \,, \\
z_L &= (1-i) \kappa_{ij} (z) \sqrt{\tau_0 (\tau + \tau_0)} \,,
\end{align}
with $\tau_0 = \sqrt{\hat q_0/E} t_0$. Due to this additional dependence on $\tau_0$ in the splitting function, the resulting distribution naturally also depends on it, i.e. $D(x,\tau) \to D(x,\tau,\tau_0)$.
\end{description}
The splitting rates have been written in terms of the color-kinematical factors $\kappa_{ij}(z)$ and (unregularised) Altarelli-Parisi splitting functions $P_{ij}(z)$ and can be found explicitly in \ref{sec:app:splitting_details}.

\subsection{Numerical results for medium-induced parton spectra}
\label{sec:dx}

\begin{figure}[t!]
\centering
	\begin{subfigure}[b]{0.48\textwidth}
	\centering
	\includegraphics[width=\textwidth]{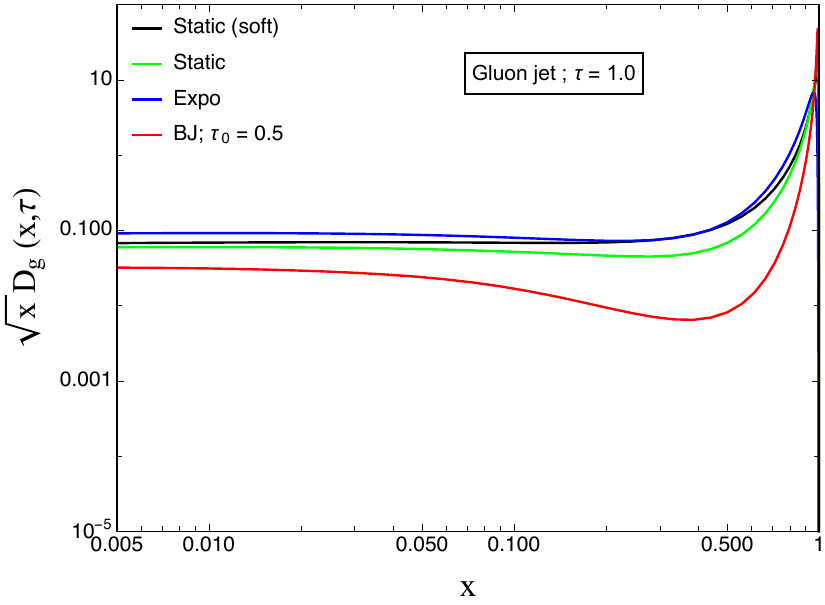}
	\includegraphics[width=\textwidth]{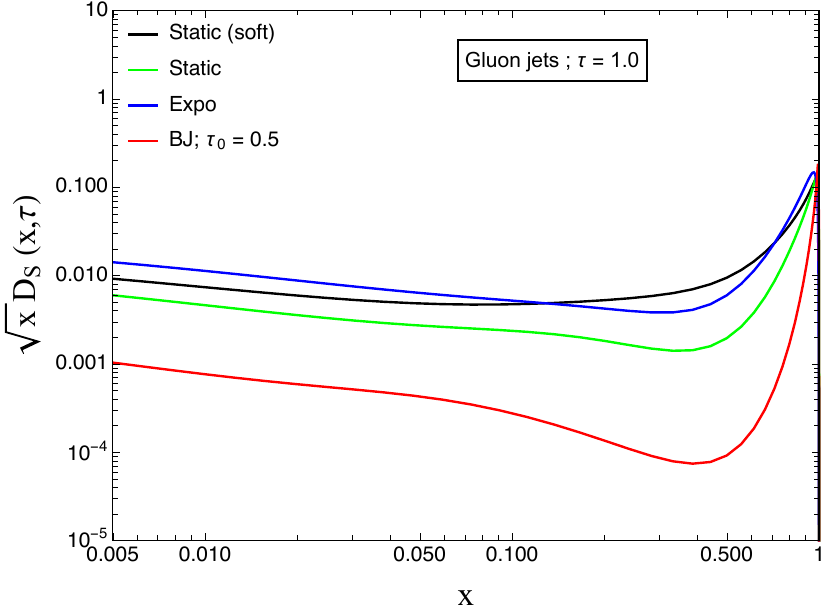}
	\caption{Gluon-initiated jets.}
	\label{fig:tau0p05_gluon}
	\end{subfigure}
	\hfill
	\begin{subfigure}[b]{0.48\textwidth}
	\centering
	\includegraphics[width=\textwidth]{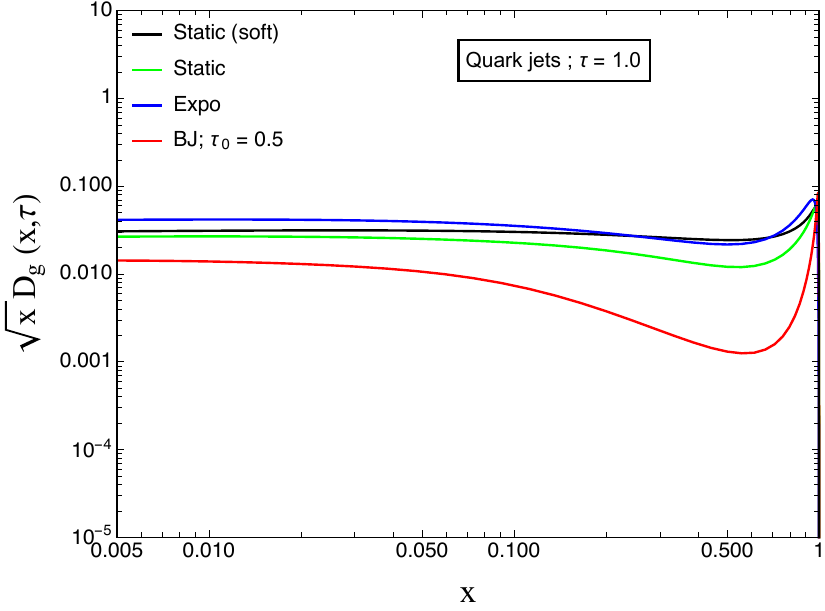}
	\includegraphics[width=\textwidth]{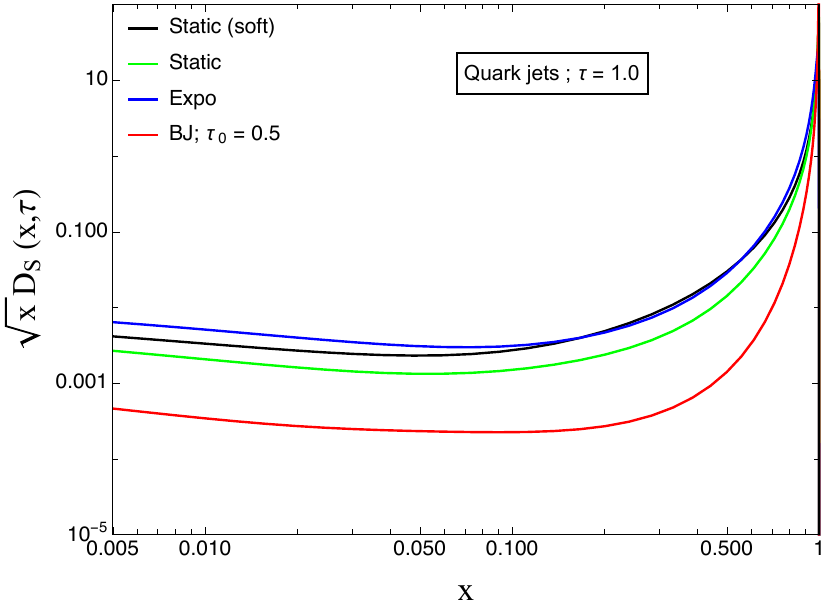}
	\caption{Quark-initiated jets.}
	\label{fig:tau0p05_quark}
	\end{subfigure}
\caption{Evolution of the quark singlet spectra $(D_S)$ and the gluon spectra $(D_g)$ for the "gluon initiated" (left panels) and the "quark initiated" jets (right panels) for a particular evolution time $\tau = 1.0$ for different medium profiles.}
\label{fig:tau0p05}
\end{figure}

We use the coupled evolution equations \eqref{eq:evol-eq-glu} and \eqref{eq:evol-eq-S} to numerically calculate the evolution of the quark and gluon spectra for both quark-initiated and gluon-initiated jets as a function of energy fraction $x$ for different evolution time $\tau$. The rates and solution for the pure gluon cascade was discussed in \cite{Adhya:2019qse}. In the current work, we solve the coupled evolution equations using finite difference methods. For future reference, the resulting distributions are evaluated for two sets of initial conditions, as follows:
\begin{itemize}
\item \textbf{gluon-initiated jets:} $D_g(x,0)=\delta(1-x)$, $D_S(x,0) = 0$.
\item \textbf{quark-initiated jets:} $D_g(x,0)=0$, $D_S(x,0) = \delta(1-x)$.
\end{itemize}
Figure~\ref{fig:tau0p05} shows the medium evolved spectra for the gluon ($D_\glu$) and the quark ($D_\S$) distributions separately for gluon-initiated (Fig.~\ref{fig:tau0p05_gluon}) and quark-initiated (Fig.~\ref{fig:tau0p05_quark}) jets as a function of $x$ evolved up to $\tau=1.0$. The evolution has been performed in media with different density profiles, i.e. performing the evolution with the different kernels in Eqs.~\eqref{eq:rate-static}, \eqref{eq:rate-expo} and \eqref{eq:rate-bjorken}, see the figure legends for details. For the Bjorken case, we have chosen $\tau_0=0.5$.

One can see that gluon spectra follow the $1/\sqrt{x}$ turbulent behavior for both quark and gluon initiated jets \emph{and} for both the static medium and the expanding medium (the static case was already discussed in \cite{Mehtar-Tani:2018zba}). However, for the quark spectra this scaling 
is only approximate for all the medium profiles including the static profile. This is mainly because the stationary behavior, i.e. $D_g(x) \sim c_1/\sqrt{x}$ and $D_S \sim c_2/\sqrt{x}$ (where $c_1$ and $c_2$ are approximately constants), has not yet had sufficient time to establish itself \cite{Mehtar-Tani:2018zba}. Note, that the speed of the evolution is slower when considering the splitting rates for finite media, as done above \cite{Adhya:2019qse}.

Outside the region $x \sim 1$, the exponential spectra have a greater amplitude than the full static spectra and similar amplitude as the static soft spectra. Spectra
for the Bjorken expanding media have the least amplitude due to a smaller rate of splittings compared to the other profiles. 
 We have shown previously \cite{Adhya:2019qse} that the ``singular'' parts of gluon spectra ($z \rightarrow 0$, $z \rightarrow 1$) for different medium profiles scale 
with the static one when using an effective quenching parameter, defined as $4\hat{q}_0$ and $4\hat{q}_0t_0/L$ for the exponential and Bjorken profiles, 
respectively. This scaling between different medium profiles holds for individual contributions discussed here as well. However, such a scaling does not work when 
employing the full evolution kernels as obtained in Ref.~\cite{Adhya:2019qse} for gluon-only cascades.

In general, the $D_\S$ and $D_\glu$ dominate the high-$x$ part of the spectra for 
quark-initiated and gluon-initiated jets, respectively. For gluon-initiated jets, 
the probability of finding a gluon in the parton cascade at high as well as 
small $x$ is significantly larger than a quark for all the cases of expanding media. 
For quark-initiated jets, the probability to find a gluon is smaller than a quark at high $x$ 
and larger at small $x$.  This is a generalization of similar observations done in Ref.~\cite{Mehtar-Tani:2018zba} for static media.

\section{Inclusive jet suppression}
\label{sec:jetsup}

  We intend to advance understanding the inclusive jet suppression in several directions in this work. 
  Baseline of this work is to quantify the impact of medium expansion on $\pt$ (Section~\ref{sec:raa}), rapidity (Section~\ref{sec:rap}), and azimuthal dependence 
(Section~\ref{sec:vtwo}) of the inclusive jet suppression in a realistic treatment of input parton spectra.
Then we quantify the sensitivity of observables on the time of the onset of the quenching (Section~\ref{sec:time}).  

  Beside this, we also quantify the impact of including nuclear parton distribution functions (nPDF), impact of including effects from early vacuum-like emissions in the plasma, 
and sensitivity on input parton spectra connected with the choice of free parameters in underlying MC modeling.
  Although we consider scenarios with expanding media, we assume a fixed in-medium path length, $L = 5$~fm, for all results except those presented in Section~\ref{sec:vtwo}.

\subsection{Transverse momentum dependence of jet suppression}
\label{sec:raa}

To quantify the suppression factor of jets in heavy-ion collisions one usually assumes a separation between the short-distance production of hard partons and their 
subsequent propagation through the medium. We define medium evolved spectra of quark and gluon cascades for quark-initiated jets (denoted by ``$\{q\}$'') and 
gluon-initiated jets (denoted by ``$\{g\}$'') as
 \begin{align}
D(x,\tau;\{q\}) &= \big[ D_g(x,\tau) + D_S(x,\tau) \big]_{D_g(x,0)=0, \,D_S(x,0)= x\delta(1-x)} \,,\\
D(x,\tau;\{g\}) &= \big[ D_g(x,\tau) + D_S(x,\tau) \big]_{D_g(x,0)=x\delta(1-x), \,D_S(x,0)=0} \,. 
\end{align}
For high-$p_T$ jets, where $\tau \lesssim 1$, one is naturally dominated by $D(x,\tau;\{ g\}) \approx D_g(x,\tau)$ and $D(x,\tau; \{q\}) \approx D_S(x,\tau)$, as discussed above.


One of the key observables quantifying inclusive jet suppression is the jet nuclear modification factor, measured by the LHC experiments \cite{Aad:2014bxa,Chatrchyan:2012raa,Aaboud:2018twu,Adam:2015ewa,Acharya:2019jyg}.
In our approach, the yield for the inclusive jet suppression can be obtained as a convolution of the $D_i(x,\tau)$ distribution with the initial parton spectra \cite{Baier:2001yt,Salgado:2003gb,Mehtar-Tani:2014yea,Adhya:2019qse}. For a given parton species $i$ that is produced in the hard cross section $\rmd \sigma^0_i/(\rmd \pT\, \rmd y)$, we write
\beq
\label{eq:medium-spectrum}
\frac{\dd \sigma^{\rm AA}_i}{\dd \pT\, \rmd y} = \int_0^1 \frac{\dd x}{x} \,D\left(x, \sqrt{x}\tau ;\{i\} \right) \left.\frac{\dd \sigma^0_{i}}{\dd \pT'\, \rmd y} \right|_{\pT' = \pT/x} \,,
\eeq
for $i=q,g$ and where the evolution time $\tau =\sqrt{\hat q_0/\pT}L$ is written in terms of the final, \emph{measured} transverse momentum $\pT$. Here, $\rmd \sigma^0/\rmd \pT$ denotes the hard spectrum for producing a parton $i$ in proton-proton collisions, or in ``vacuum''. For the Bjorken case, where the distribution has an explicit dependence on the initial time $\tau_0$, we approximate $D(x,\sqrt{x}\tau, \sqrt{x}\tau_0) \approx D(x,\sqrt{x} \tau, \tau_0)$ \cite{Adhya:2019qse} since $\tau_0$ is already very small and the integral \eqref{eq:medium-spectrum} is dominated by $x\sim 1$. The quenching factor for parton species $i$, ${\cal Q}_i(\pT)$ , defined as the ratio of partonic medium and vacuum spectra, is then
\beq
{\cal Q}_i(\pT)\equiv \frac{\dd \sigma^{\rm AA}_i \big/(\dd \pT\, \rmd y)}{\dd \sigma^0_i \big/(\dd \pT\, \rmd y)}  \,,
\eeq
measured at the same final $\pT$ and the same rapidity $y$ (or rapidity range).
If we approximate the initial parton spectrum by a power law $\dd \sigma^0/\dd \pT \propto \pT^{-n}$, with fixed power index $n=$const., the quenching factor is then a $(n-1)^{st}$ moment of the in-medium parton distribution, i.e. ${\cal Q}_i = \int_0^1 \rmd x\, x^{n-1}D(x,\sqrt{x}\tau; \{i\})$. 

In this study, we go beyond this simple approximation and extend the power law into a more general form
\beq
\label{eq:MPL}
\frac{\dd \sigma^0_i}{\dd \pT\, \rmd y} = a_{i} \left(\frac{\pTz}{ \pT}\right)^{n_i(\pT,y)} \,,
\eeq
where $n_i(\pT,y) = n_{i,0}-\beta_i \log(\pT/\pTz)-\gamma_i \log^2(\pT/\pTz)-\delta_i \log^3(\pT/\pTz)$
which provides a good characterization of the initial parton spectra \cite{Spousta:2015fca}. Here $\beta_i$, $\gamma_i$, and $\delta_i$ are additional free parameters to be fitted separately for quark- and gluon-initiated jets. The rapidity dependence is parameterized via the coefficients $n_{i,0}$, $\beta_i$, $\gamma_i$ and $\delta_i$ that are extracted for each rapidity selection. 
  
In this case, the 
   quenching factor for a given parton species $i$, is written as
\beq
\label{eq:suppression-factor-2}
{\cal Q}_i(\pT) = \int_0^1 \dd x \,x^{n_i(p_T,y)-1} D(x, \sqrt{x} \tau;\{i\}) \,.
\eeq
  This factor describes the effect that energy loss has on the spectrum of a single parton. However, a jet is a multi-parton state due to the large phase space for 
  radiation, and at high-$p_T$ many of these emissions will take place with momentum scales that are much higher than the typical medium scales attainable via 
  multiple scattering in the medium \cite{Mehtar-Tani:2017web}. Every emission of this type corresponds to a resolved color charge that contributes toward the total 
  energy loss of the full jet. The resummation of such emissions in an expanding medium is derived in \ref{sec:fulljet-quenching}, see also \cite{Caucal:2020uic} for 
  a related discussion. Implementation of these vacuum-like emissions (VLE) can be turn on and off allowing to assess their impact on measurable quantities as discussed further in this section. 

The full nuclear modification factor contains the contributions from both quarks and gluons to the final jet spectra, and reads
\beq
\Raa (\pT) = \frac{\sum_{i=q,g} \frac{\dd \sigma^{\rm AA}_i}{\dd \pT} }{\sum_{i=q,g} \frac{\dd \sigma^0_i}{\dd \pT} } \,.
\eeq
Finally, the combined nuclear modification factor $\Raa$ reads 
\beq
\label{eq:combRaa}
\Raa = f_q(\pT,y,R)\,{\cal Q}_q(\pT,R) + f_g(\pT,y,R) \, {\cal Q}_g(\pT,R) \,,
\eeq
where 
\beq
\label{eq:fq}
       f_i(\pT,y,R) = \frac{\sigma_i^{\rm 0,med}(\pT,y,R)}{\sigma_q^{\rm 0}(\pT,y,R) + \sigma_g^{\rm 0}(\pT,y,R)} \,.
\eeq
Above, we have explicitly included the dependence of jet cone-size $R$ on the generated spectra and the quenching factors.
Here $\sigma_i^{\rm 0,med}$ is the vacuum cross-section for the hard production which takes into account nPDF effects. If these are to be neglected, this factor is replaced by $\sigma_i^{\rm 0}$.

Parameters of the extended power law description of input parton spectra were obtained from fits of jet spectra from PYTHIA8 
\cite{Sjostrand:2007gs}.
  In general, three versions of spectra were used to study the sensitivity of observables on modeling the input parton spectra: 1) spectra based on PYTHIA8.185 with 
AU2 tune \cite{ATLAS:2012uec} and CT10 PDFs \cite{Lai:2010vv} ( configuration labeled ``vacuum 1''), 2) spectra based on default version of PYTHIA 8.306
(configuration labeled ``vacuum 2''), and 3) spectra based on default version of PYTHIA 8.306 with nPDFs from EPS09LO \cite{Eskola:2009uj} (configuration labeled ``nPDF'').
  The power-law parameters for the three configurations are summarized in Tab.~\ref{tablevac2nPDF} and Tab.~\ref{tab:inputs}. Jet spectra were obtained by running FastJet \cite{Cacciari:2011ma} using the anti-$k_{t}$ algorithm \cite{Cacciari:2008gp} with resolution parameter $R=0.4$ for the center of mass energy of $\sqrt{s_\mathrm{NN}} = 5.02$~TeV. Jet spectra were reconstructed 
separately for several rapidity selections specified in Tab.~\ref{tablevac2nPDF}  and used later in the study. Distributions calculated inclusively in rapidity were
obtained for rapidity selection of $|y|<2.8$. The jets were matched to one of two outgoing partons from the leading order hard-scattering process by choosing the 
parton with the smallest angular distance, $R = \sqrt{ \Delta \phi^2 + \Delta \eta^2}$, and the flavor of the jet was assigned to be that of the matched parton. 
Furthermore, for all the numerical results presented below we set $\alpha_s = 0.1$. The expressions for quenching factor, jet $R_{AA}$, and flavor fraction in the configuration with no nPDF effects and no VLE as well as power-law parameters for simple ``vacuum 1'' configuration are provided in \ref{app:raa}.

\begin{table}[]
\centering
\begin{tabular}{|c|c|c|c|c|c|c|}
\hline
parameters & \begin{tabular}[c]{@{}c@{}}$|y| < 0.3$\\  (vac 2)\end{tabular} & \begin{tabular}[c]{@{}c@{}}$|y| < 0.3$ \\ (nPDF)\end{tabular} & \begin{tabular}[c]{@{}c@{}}$|y| < 2.8$ \\ (vac 2)\end{tabular} & \begin{tabular}[c]{@{}c@{}}$|y| < 2.8$ \\ (nPDF)\end{tabular} & \begin{tabular}[c]{@{}c@{}}$2.1 < |y| < 2.8$ \\ (vac 2)\end{tabular} & \begin{tabular}[c]{@{}c@{}}$2.1 < |y| < 2.8$ \\ (nPDF)\end{tabular} \\ \hline
$n_{g,0}$ & 2.51 & 4.39 & 4.39 & 5.43 & 5.55 & 4.49 \\ \hline
$n_{q,0}$ & 4.31 & 3.38 & 3.91 & 4.22 & 4.97 & 3.35 \\ \hline
$\beta_{g}$ & 2.69 & 1.16 & 1.38 & 0.55 & 1.68 & 2.61 \\ \hline
$\beta_{q}$ & 0.91 & 1.67 & 1.20 & 1.04 & 1.52 & 3.30 \\ \hline
$\gamma_{g}$ & -1.06 & -0.52 & -0.61 & -0.20 & -1.09 & -1.56 \\ \hline
$\gamma_{q}$ & -0.46 & -0.71 & -0.47 & -0.43 & -0.86 & -1.93 \\ \hline
$\delta_{g}$ & 0.17 & 0.12 & 0.13 & 0.10 & 0.63 & 0.58 \\ \hline
$\delta_{q}$ & 0.11 & 0.13 & 0.09 & 0.10 & 0.50 & 0.61 \\ \hline
$a_{g}$ (nb) & $4.23 . 10^{-4}$ & $1.51 . 10^{-4}$ & $1.1 . 10^{-3}$ & $8.27 . 10^{-5}$ & $2.10 . 10^{-5}$ & $8.82 . 10^{-5}$ \\ \hline
$a_{q}$ (nb) & $2.99 . 10^{-5}$ & $6.47 . 10^{-5}$ & $8.33 . 10^{-4}$ & $3.24 . 10^{-4}$ & $1.35 . 10^{-5}$ & $4.76 . 10^{-5}$ \\ \hline
$p_{T,  g, 0}$ (GeV) & 27.28 & 39.18 & 37.88 & 65.58 & 59.26 & 44.99 \\ \hline
$p_{T, q, 0}$ (GeV) & 44.34 & 35.98 & 33.39 & 41.82 & 59.20 & 43.71 \\ \hline
\end{tabular}
\caption{List of parameters for the configurations ``vacuum 2" and "nPDF".}
\label{tablevac2nPDF}
\end{table}

\begin{figure}
\centering
\includegraphics[scale=0.5]{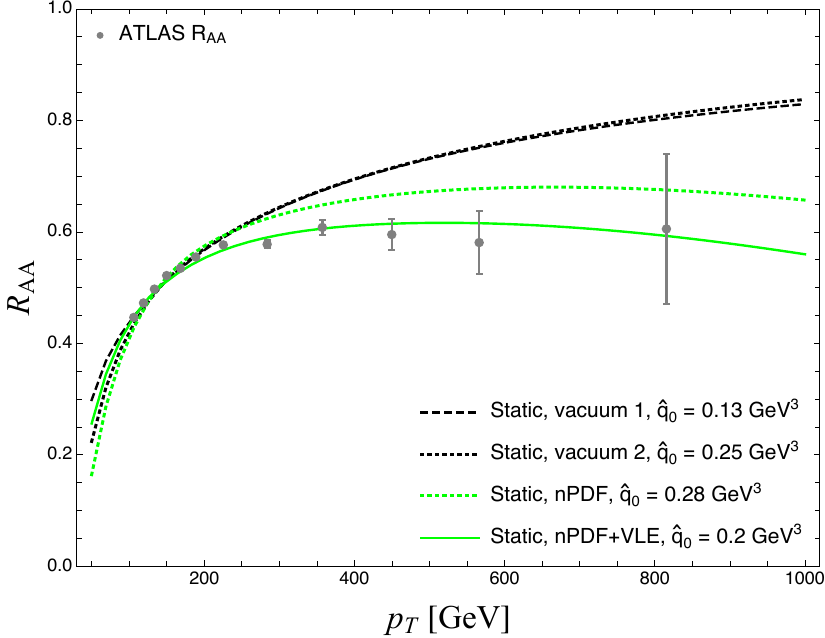}~~~~~~\includegraphics[scale=0.5]{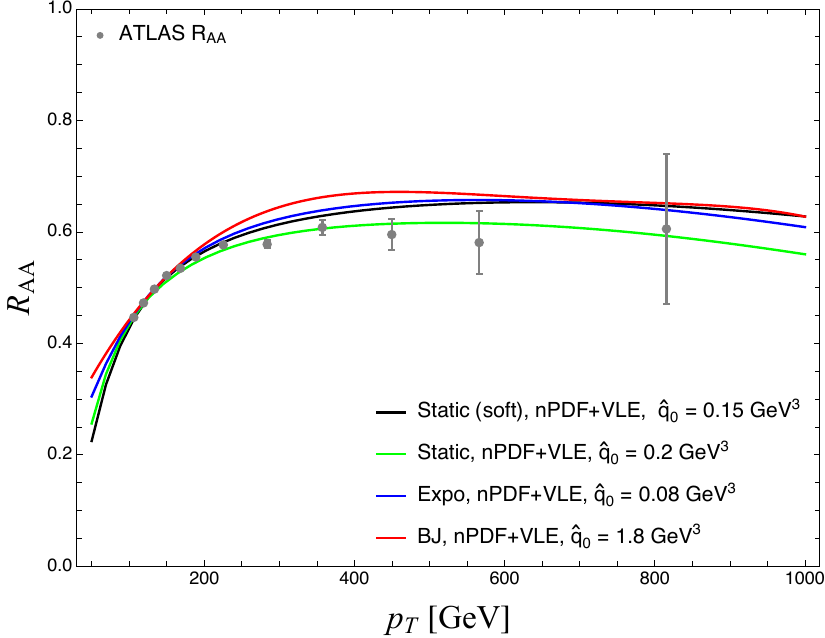}
\caption{
  Left: jet $\Raa$ for static medium in four configurations: two version of vacuum input parton spectra (``vacuum 1,2''), spectra with nPDF effects (``nPDF'') and spectra with
  nPDF effects and vacuum-like emissions included (``nPDF+VLE''). Right: jet $\Raa$ for ``nPDF+VLE'' configuration for four medium profiles: 
  static soft (black), static (green), exponential (blue), and Bjorken with $t_0=0.1$~fm (red)
  The $\qhat_0$ values were optimized for each profile to reproduce the data (in gray) \cite{Aaboud:2018twu}. 
}
\label{fig:raa1qgini}
\end{figure}
  Summary of calculated $\pt$ dependence of nuclear modification factor $\Raa$ is plotted in Fig.~\ref{fig:raa1qgini} along with experimental data on
measured anti-$k_T$ jets with $R=0.4$ \cite{Aaboud:2018twu}. For each configuration, we use value of $\qhat_0$ obtained from the $\chi^2$ minimization of the difference 
in the nuclear modification factor between the data and the theory. We fixed the parameter $L$ to 5~fm for all the medium profiles. 

  Left panel of Fig.~\ref{fig:raa1qgini} compares results for static medium for four configurations: two versions of input parton spectra (``vacuum 1'', 
``vacuum 2''), input parton spectra with nPDF effects included (``nPDF''), and input parton spectra with nPDF effects and vacuum-like emissions included 
(``nPDF+VLE''). One can see that shape of the $\Raa$ can be fully described only after implementing nPDF effects. Comparison of ``vacuum 1'' and ``vacuum 2'' 
configurations shows minimal sensitivity of the shape of $\Raa(\pt)$ on PYTHIA version of in-vacuum input parton spectra. 
Comparison of ``nPDF'' and ``nPDF+VLE'' configuration implies that adding vacuum-like emissions to the calculation has an 
important impact on both the shape of $\Raa$ and its overall normalization which has an impact on the extracted values of $\qhat_0$.
  
  Right panel of Fig.~\ref{fig:raa1qgini} shows $\Raa$ for four types of medium expansion in the configuration ``nPDF+VLE''. One can see that the difference in the 
shape and magnitude of $\Raa$ among different medium profiles is rather small, less than 10\% for $\pt > 100$~GeV. The obtained $\qhat_0$ values for realistic quark 
and gluon initial spectra in the configuration of nPDF+VLE are tabulated together with the values obtained for a gluon-only calculation with a simple power-law 
spectrum and no nPDF and VLE effects as published in Ref.~\cite{Adhya:2019qse} in Table~\ref{tab:qhat}.
  The extracted value of the jet quenching parameter differs significantly between the simple configuration 
and the more realistic treatment.
  The driving factor of this difference is the use of both the quark and gluon degrees of freedom and improved description of input parton spectra.  
  This illustrates and quantifies a simple fact that the value of jet quenching parameter has an important dependence on the level of completeness of the jet energy loss description and 
approximations used.
  Since the $\Raa$ distributions do not scale when using the above quoted $\hat{q}_0$ values, we confirm the breakdown of the ``effective'' scaling laws for $\Raa$, 
which was discussed in Ref.~\cite{Adhya:2019qse}, also for more realistic treatment presented in this paper.

The results in Fig.~\ref{fig:raa1qgini} shows that the single-parton description of the jet $\Raa$ used here describes the shape of the experimental data in the 
$\pt$ region below $\sim 250$~GeV. On the contrary, at $\pt \gtrsim 250$~GeV the calculated $\Raa$ is steeper compared to the data. This was already observed in 
Ref.~\cite{Adhya:2019qse} for the gluon-only calculation and the more complete description of jet energy loss used here does not lead to an improvement in the 
description of the data. 
Comparison of ``nPDF+VLE'' configuration with the data suggests that the flatting of $\Raa$ at high-$\pt$ is due to the combination of nuclear PDFs, that produce a suppression at high-$\pT$ due to the EMC effect (see also \cite{Pablos:2019ngg,Huss:2020dwe}), and the onset of the quenching of multiple resolved partons in a jet \cite{CasalderreySolana:2012ef}. These appear due to 
hard vacuum-like splittings inside the medium \cite{Mehtar-Tani:2017web}, see also \cite{Caucal:2018dla}. A resummation of such fluctuations leads to stronger 
suppression at high-$\pT$ due to enhanced energy loss as discussed in Refs.~\cite{Rajagopal:2016uip,Casalderrey-Solana:2019ubu,Mehtar-Tani:2021fud,Takacs:2021bpv}.

The inability to describe the full $\pt$ region of measured $\Raa$ in the configuration without the nPDF+VLE effects is in contrast with parametric modeling 
presented in Refs.~\cite{Spousta:2015fca,Aaboud:2018twu}, where the use of improved description of input parton spectra and use of appropriate quark and gluon mixing 
was shown to lead to a good description of the $\Raa$ over a large $\pt$ range under the assumption of $\pt$-dependence of the energy loss which was close to 
$\sqrt{\pt}$-dependence. Within the BDMPS, the $\sqrt{\pt}$-dependence of the lost energy holds at low-$\pt$, but does not hold at high-$\pt$ \cite{Baier:2001yt}, 
where the single-scattering contribution to the medium-induced energy loss dominates in calculations \cite{Wiedemann:2000za}, see also 
\cite{Mehtar-Tani:2021fud,Takacs:2021bpv}, and where nPDF effects play a role. The power of 0.55 used in parametric modeling of Refs.~\cite{Spousta:2015fca} may 
effectively account for these phenomena, however, without providing ability to distinguish them.

\begin{table}[]
\centering
\begin{tabular}{|l|l|l|l|c|}
\hline
\multicolumn{1}{|c|}{\multirow{2}{*}{\begin{tabular}[c]{@{}c@{}}Quenching\\ parameter ($\hat{q}$)\end{tabular}}} & \multicolumn{1}{c|}{\multirow{2}{*}{Static (soft)}} & \multirow{2}{*}{Static} & \multirow{2}{*}{Expo} & \multicolumn{1}{c|}{Bjorken}                                         \\ 
\multicolumn{1}{|c|}{}                                                                                           & \multicolumn{1}{c|}{}                               &                         &                       & $t_0 = 0.1$~fm \\ \hline
$\hat{q}_0$ (nPDF+VLE) [GeV$^3$]                                                                                     & 0.15                                                & 0.2                   & 0.08                 &  1.8              \\ \hline
$\hat{q}_0$ (gluon-only) [GeV$^3$]                                                                               & 0.20                                                & 0.2                     & 0.09                 & 2.6       \\ \hline
\end{tabular}
\caption{The jet quenching parameter for different media profiles obtained from fitting the calculation to the data from Ref.~\cite{Aaboud:2018twu}. 
  Full description of the input parton spectra (first line) can be compared with results using gluon-only cascades and power-law description of input parton spectra 
(second line), published in Ref.~\cite{Adhya:2019qse}.}
\label{tab:qhat}
\end{table}

~

\subsection{Initial-time dependence of jet suppression}
\label{sec:time}

\begin{figure}
\includegraphics[width=0.45\textwidth]{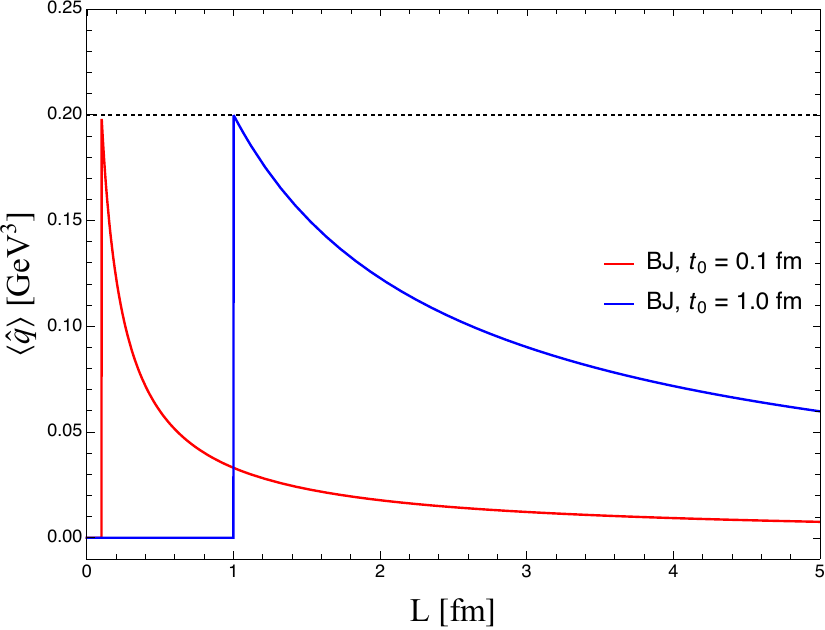}~~~~~
\includegraphics[width=0.45\textwidth]{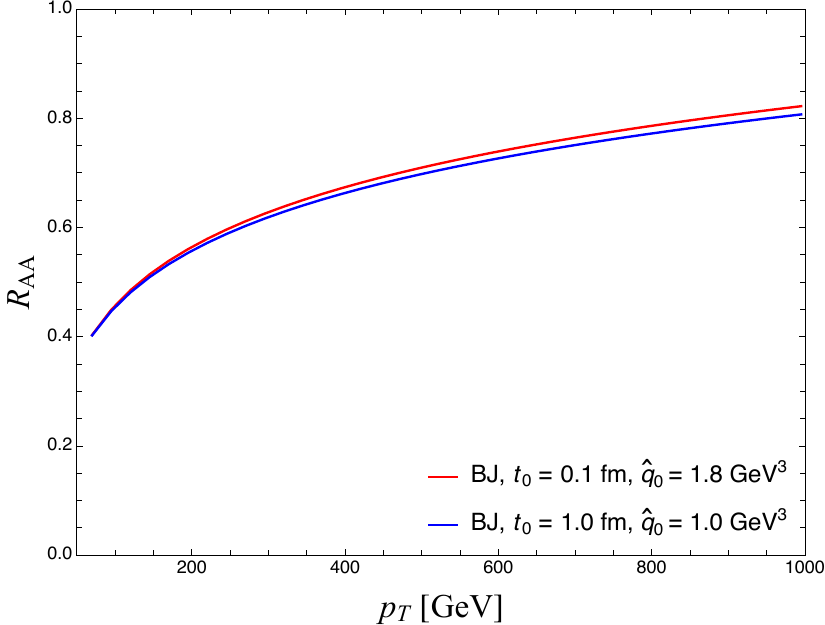}
\caption{Left: The $\qhat$ as a function of the length of medium, $L$, for Bjorken medium for two values of initial time of quenching, $t_0=0.1$~fm and $t_0=1.0$~fm. 
Right: Jet suppression factor $\Raa$ as a function of $\pt$ for Bjorken medium with $L = 5$~fm and with $\qhat_0$ values optimized to reproduce the experimental data \cite{Aaboud:2018twu}. Note that we have not included nPDF effects in this particular figure.
}
\label{fig:BjRaa}
\end{figure}

  The value of $t_0$ in the Bjorken medium quantifies the time during which no interaction with the medium occurs. This reflects a situation with a finite 
time needed for the onset of the quenching interactions in the medium. In this section, we discuss the impact of the choice of $t_0$ on the observable jet suppression 
by comparing two different starting times of the quenching in the Bjorken medium, namely $t_0 = 0.1$~fm and $t_0 = 1.0$~fm, in the most simple configuration with no nPDF effects and vacuum-like emissions included.
  Left panel of Fig.~\ref{fig:BjRaa} shows a comparison of the $\langle \hat{q} \rangle$ for the Bjorken media with these two choices of $t_0$ evaluated 
as a function of $L$ calculated as
\beq
\label{eq:qhat-bjorken}
\langle \hat{q} \rangle = \frac{2}{(L-t_0)^2}\int_{t_0}^{L}\rmd t\, (t-t_0) \hat{q}(t)  \,,
\eeq
where $\hat q(t) = \hat q_0 t_0/t$ for the Bjorken scenario. One can see that delaying the onset of quenching leads to a larger average value of $\hat{q}$ at $L > t_0$ compared to an earlier time of the quenching. In general, the later the quenching sets in, the larger the $\langle 
\hat{q} \rangle$ is, which makes the medium more opaque at the later stages of the evolution.

We use again the $\chi^2$ minimization procedure to find the $\qhat_0$ value which minimizes the difference in the $\Raa$ between the data and the theory prediction 
for the two $t_0$ values with medium length fixed to $L = 5$~fm. The results  are shown in the right panel of Fig.~\ref{fig:BjRaa} (note that we have left out nPDF effects in this particular figure). For $t_0=0.1$ 
and $1.0$~fm, the $\qhat_0$ is found to be approximately 1.8~GeV$^3$ and 1.0~GeV$^3$, respectively. That is for ten times larger initial time, the $\qhat_0$ has to 
be roughly two times smaller to obtain the same $\Raa$. While the choice of $t_0$ can be somehow absorbed into an adjusted value of $\hat q_0$, we will see below 
that it affects other observables. In the next two sections, we use optimized $\qhat_0$ values, and their respective $t_0$ parameters, to calculate more 
differential observables derived from $\Raa$ which characterize the inclusive jet suppression and quantify how these settings affect them.

\subsection{Rapidity dependence of jet suppression}
\label{sec:rap}

\begin{figure}
\centering
	\begin{subfigure}[b]{0.31\textwidth}
		\centering
\includegraphics[width=\textwidth]{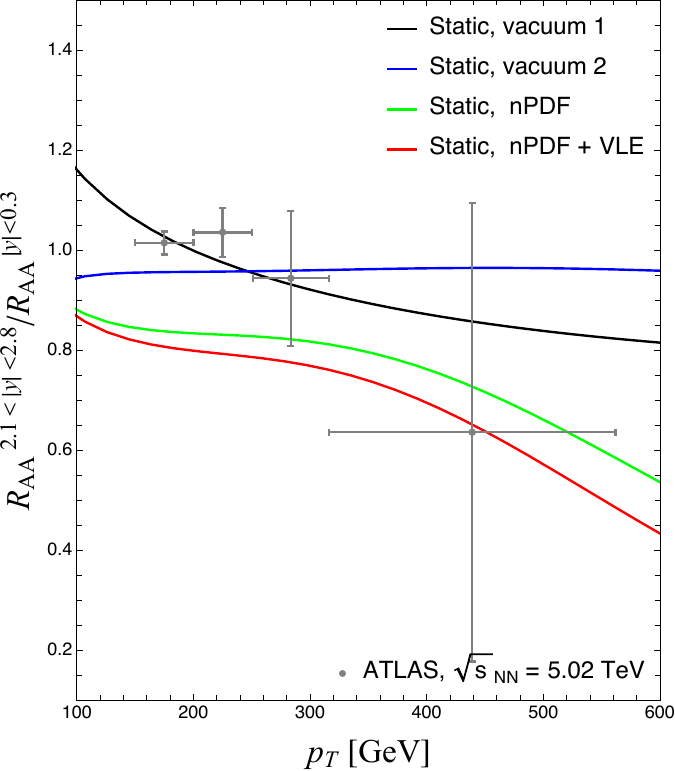}
		\caption{$\pT$ dependence.}
		\label{fig:Qratiorapsopti-pT}
	\end{subfigure}
	\hfill
	\begin{subfigure}[b]{0.66\textwidth}
		\centering
\includegraphics[width=0.46\textwidth]{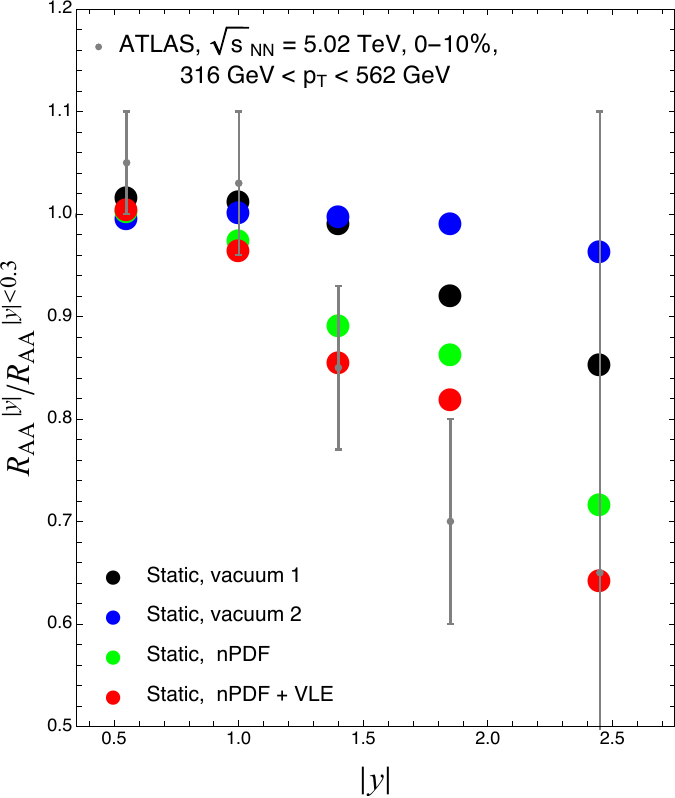}
		~~~~\includegraphics[width=0.46\textwidth]{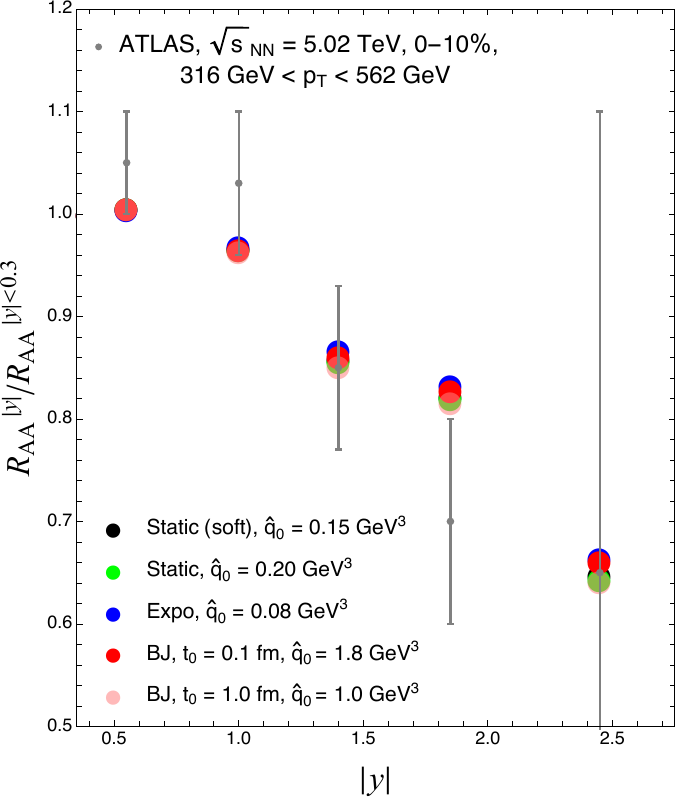}
		\caption{$y$ dependence.}
		\label{fig:Qratiorapsopti-y}
	\end{subfigure}
\caption{
  Left: The ratio of $\Raa$ in $2.1<|y|<2.8$ and $\Raa$ in $|y|<0.3$ as a function of jet $\pt$. Middle and Right: The ratio of $\Raa$ in different $|y|$ bins and 
$\Raa$ in $|y|<0.3$ for $\pt = 316-512$~GeV. Black, blue, green, and red color encodes static medium for ``vacuum 1'', ``vacuum 2'', static medium with nPDF effects, 
and with nPDF+VLE effects, respectively, in the left and middle panels. Black, green, blue, red and pink color encodes static soft, static, exponential, and two 
Bjorken medium profiles, respectively, in the right panel.
  The ATLAS data taken from Ref.~\cite{Aaboud:2018twu}. 
 }
\label{fig:Qratiorapsopti}
\end{figure}

Following the measurement done in Ref.~\cite{Aaboud:2018twu}, we study the rapidity dependence of the jet suppression. The optimized $\qhat_0$ values from 
Secs.~\ref{sec:raa} and \ref{sec:time} 
  are used to calculate the rapidity dependent $\Raa$. The results are 
presented in Fig.~\ref{fig:Qratiorapsopti} in terms of the ratio of the $\Raa$ in a given rapidity region to the $\Raa$ in central rapidity region ($|y|<0.3$). 
Figure~\ref{fig:Qratiorapsopti-pT} shows the ratio of $\Raa$ in $2.1 < |y| < 2.8$ and $|y|<0.3$ as a function of $\pT$ for the static medium for four 
configurations: two different versions of input parton spectra (``vacuum 1'', ``vacuum 2''), input parton spectra with nPDF effects included (``nPDF''), and both 
nPDF and VLE effects included (``nPDF+VLE''). One can see that the resulting ratio differs significantly among different configurations. 

Left panel of 
Fig.~\ref{fig:Qratiorapsopti-y} detail the rapidity dependence of $\Raa$ ratio for a fixed $\pT$, namely 316 GeV $<\pT<$ 562 GeV. One can see that inclusion of nPDF 
and VLE effects leads to qualitative agreement between trends seen in the data and calculations. At the same time a trend of decreasing $\Raa$ ratio is partially 
included in the second version of PYTHIA used in this study as well. 
Here, the differences between the different PYTHIA spectra should be taken as an indication of an overall theoretical uncertainty of the rapidity behavior.

Right panel of Fig.~\ref{fig:Qratiorapsopti-y} shows the same quantity as the middle panel evaluated for default version of input parton spectra configuration for 
five medium profiles. 
Results for static, exponential, and two Bjorken profiles discussed in Sec.~\ref{sec:time} are found to differ by less then 10\% for all the rapidity bins (and 
also for all the $\pt$ values which is not shown in the figure). 

  The fact that the rapidity dependence of the inclusive jet suppression is very similar for different medium profiles
  implies that 
the origin of the rapidity dependence is somewhat universal. To check this, we have calculated the ratio of $\Raa$ for gluon-only evolution. When evaluated as a 
function of $|y|$ we see the same trends as in the default case that includes both quark and gluon fragmentation. This implies that the rapidity dependence is 
in part a consequence of the change in the steepness of input parton spectra. However, both the nPDF and VLE effects 
  have clearly larger impact on this observable quantity.

\subsection{Path-length dependence of jet suppression and jet $v_2$}
\label{sec:vtwo}

The path-length dependence of jet suppression can be studied using the jet $v_2$ evaluated in a simple approximation \cite{Zigic:2018smz} as

\begin{equation}
\label{eq:v2}
v_2 = \frac{1}{2} \frac{\Raa(L^{in}) - \Raa(L^{out})}{\Raa(L^{in}) + \Raa(L^{out})},
\end{equation}
 where $L^{in}$ and $L^{out}$ are path lengths in the direction of the event plane and in the direction perpendicular to the event plane, respectively. Values of 
$L^{in}$ and $L^{out}$ were obtained from a hard sphere model where vertices of hard processes and directions of back-to-back initial partons in the azimuthal plane were 
randomly generated in the overlapping region of two spheres. The mapping between the impact parameter and $N_{coll}$ values is done using a Glauber 
model~\cite{Loizides:2017ack}. It should be stressed that this approach represents a tool for basic investigation of the relation between the medium expansion and the 
jet $v_2$, rather than a modeling with a full description of the problem. The values $L^{in}$ and $L^{out}$ for different centralities are listed in Tab.~\ref{tab:hardsphereparams}. 
  As the impact of nPDFs and vacuum-like emissions has no intrinsic azimuthal dependence and it cancels in the ratio (\ref{eq:v2}), we show the results only for the 
  default input parton configuration.

\begin{figure}
\centering
	\begin{subfigure}[b]{0.48\textwidth}
	\centering
	\includegraphics[width=\textwidth]{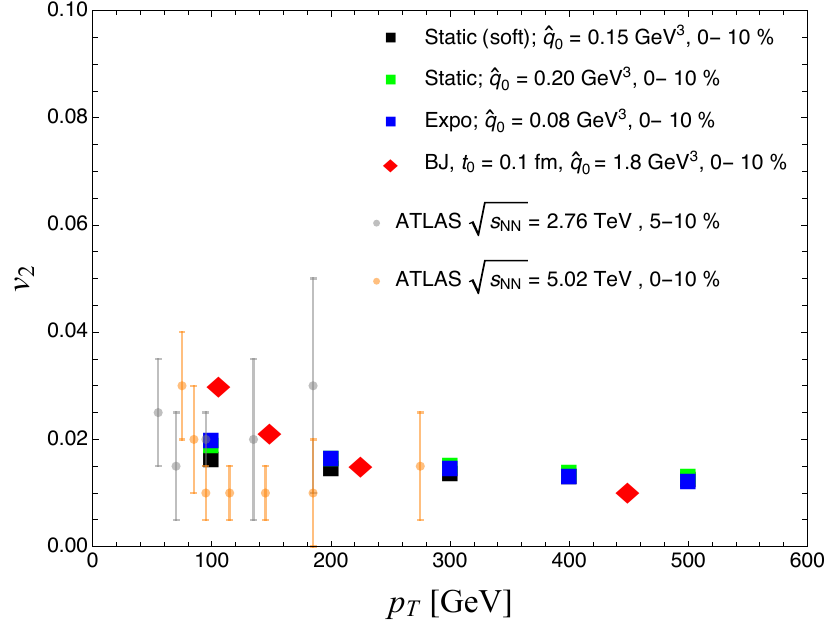}
	\includegraphics[width=\textwidth]{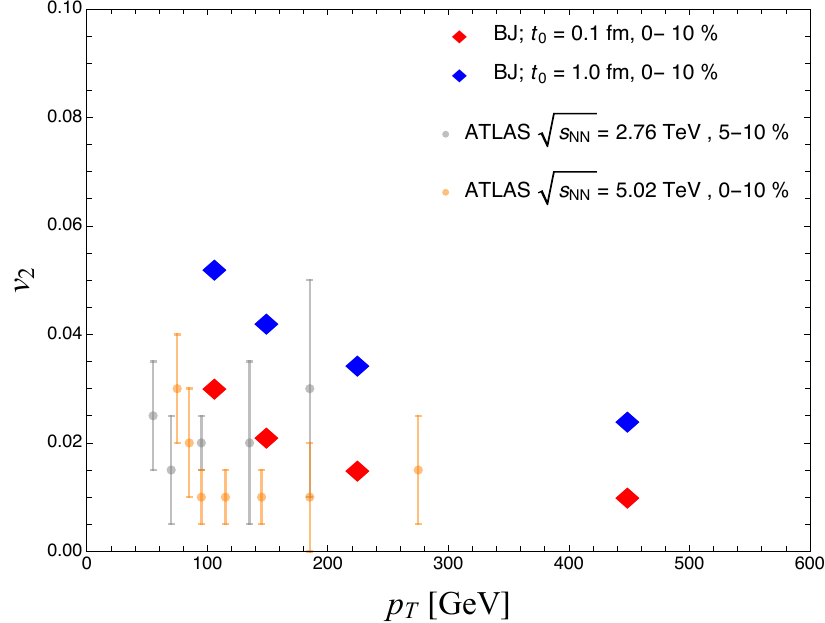}
	\caption{$\pT$ dependence.}
	\label{fig:v2allprofs-pT}
	\end{subfigure}
	\hfill
	\begin{subfigure}[b]{0.48\textwidth}
	\centering
	\includegraphics[width=\textwidth]{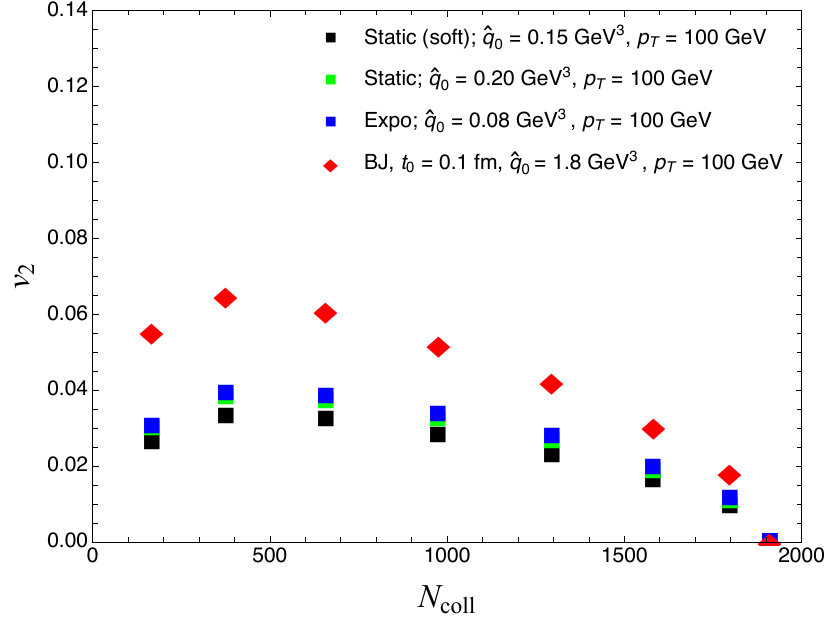}
	\includegraphics[width=\textwidth]{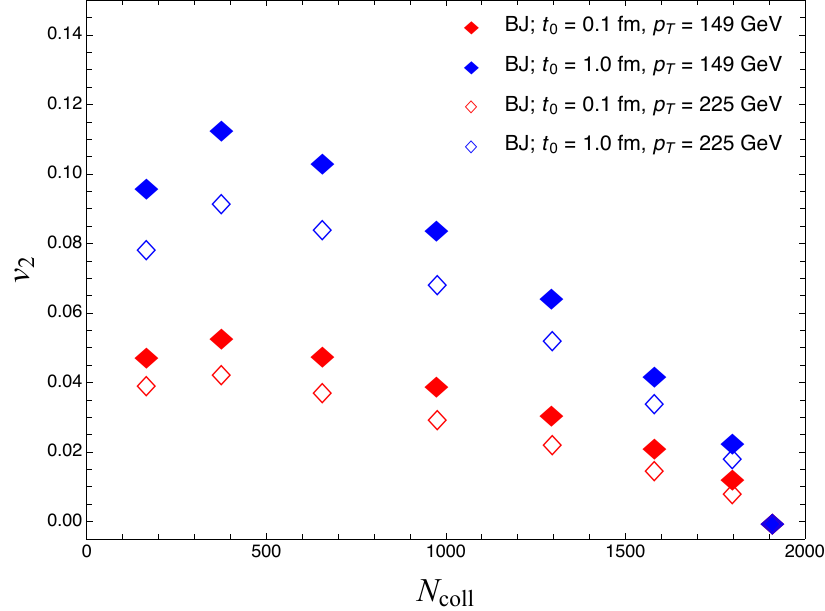}
	\caption{Centrality dependence.}
	\label{fig:v2allprofs-Ncoll}
	\end{subfigure}
\caption{
  The jet $v_2$ elliptic flow for the static and expanding profiles with optimized $\hat{q_0}$ as a function of impact parameter $\pT$ (left) and $N_{coll}$ right. 
Upper panels show the jet $v_2$ for static medium, exponential medium, and Bjorken medium with $t_0 = 0.1$~fm. Lower panels show results for Bjorken media for 
$t_0=0.1$~fm and $t_0=1.0$~fm.
}
\label{fig:v2allprofs}
\end{figure}

The upper left panel of Fig.~\ref{fig:v2allprofs} shows the jet $v_2$ evaluated as a function of $\pT$ for central collisions ($N_{coll} = 1580$) along with the 
result of measurements from Refs.~\cite{Aad:2013sla,ATLAS:2020qxc}. One can see that the magnitude and the trends of the jet $v_2$ distribution obtained from the 
calculation match those seen in the data. The upper right panel of Fig.~\ref{fig:v2allprofs} shows the jet $v_2$ evaluated as a function of $N_{coll}$. Both $\pT$ and 
$N_{coll}$ dependence of jet $v_2$ show rather small differences among static, static soft, and exponential medium. The results for Bjorken medium with $t_0=0.1$~fm 
differ significantly from results for other medium profiles.
  Even larger difference can be seen for the Bjorken medium with two different choices of $t_0$ shown in lower panels of Fig.~\ref{fig:v2allprofs}.
  These observations have similar implications as those discussed in the case of the rapidity dependence of $\Raa$: the impact of the medium expansion can be largely 
scaled out by a suitable choice of $\qhat_0$, which is however not the case for the starting time of the quenching; the jet $v_2$ remains sensitive to choice of $t_0$. 
This is in agreement with findings of the sensitivity of $v_2$ on $t_0$ published in Ref.~\cite{Andres:2019eus} which was done in more complex modeling of the 
collision geometry, but less complex modeling of the medium induced showering than in this paper.

\begin{table}[]
\centering
\begin{tabular}{|c||c|c|c|c|c|c|c|c|c|} \hline
$N_{coll}$  &  1910  &  1800  &  1580  &  1300  &  975  &  660  &  380  &  170  &  70  \\ \hline 
$L_{in}$    &  5.6   &  4.8   &  4.3   &  3.6   & 3.0   &  2.4  &  1.8  &  1.1  &  0.5 \\ \hline
$L_{out}$   &  5.6   &  5.5   &  5.2   &  5.0   & 4.7   &  4.3  &  3.7  &  2.7  &  2.1 \\ \hline
\end{tabular}%
\caption{The length parameters $L_{in}$ and $L_{out}$ from the hard sphere model for ten values of number of binary collisions, $N_{coll}$.
}
\label{tab:hardsphereparams}
\end{table}

\section{Summary and Conclusions}
\label{sec:summary}

Both quark and gluon degrees of freedom were used to calculate the evolution of partonic cascades in expanding media (static, exponential, and Bjorken). This together
with an improved description of input parton spectra makes the modeling of medium induced radiation in the limit of multiple soft scattering more realistic compared to
gluon-only cascades and power-law modeling of input parton spectra.

Using numerical estimates of medium-induced parton spectra, the nuclear modification factor $\Raa$ of jets at a fixed cone-size $R=0.4$ was calculated and jet quenching 
parameter, $\qhat$, was extracted by comparing the calculations with the data. The differences between extracted $\qhat$ in the improved description and in the simple, 
gluon-only scenario \cite{Adhya:2019qse}, were found to be significant (e.g. for Bjorken medium with $t_0 = 0.1$~fm the $\qhat_0$ value is by $30\%$ smaller in the 
improved description compared the simple gluon-only case).  At the same time, the shape of the $\Raa$ is not modified significantly in the improved description 
compared to the simple scenario. 
  Only when including effects from nuclear parton distribution functions and vacuum-like emissions, the shape of the calculated $\Raa$ starts matching the measured 
$\pt$ dependence observed in the data.

The rapidity dependence of the jet suppression is studied and it is found that the decrease of the $\Raa$ seen in the forward region by ATLAS \cite{Aaboud:2018twu} is 
consequence of multiple effects, namely the change in the steepness of input parton spectra and impact of effects from nuclear parton distribution functions and vacuum-like emissions. 
  When comparing results for different medium profiles we conclude that the rapidity 
dependence of the inclusive jet suppression is not very sensitive to the way how the medium expands.
  Further, the jet $v_2$ is studied in an approximation of hard-sphere geometry. 
  It is found that the differences in the jet $v_2$ between full calculation of static medium, static soft, and exponential medium
are within 15\%. On the contrary, 
jet $v_2$ differs in maximum roughly by a factor of two between Bjorken scenario with $t_0=0.1$~fm and $t_0=1.0$~fm. 
 %
 %
  These findings indicate that the simultaneous study of jet $v_2$ and rapidity dependence of the jet 
production may help disentangling various effects entering the observed jet quenching. 

The presented study, which quantifies the inclusive jet suppression differentially in the full set of basic kinematic quantities -- transverse momentum, rapidity, and the azimuth, represents one more step towards realistic modeling of the parton energy loss. It should contribute to the effort of disentangling the impact of different effects on measured quantities characterizing inclusive jet suppression.
%
\begin{acknowledgements}
  We would like to thank the referee of our paper, whose comments and questions helped
improve the presentation of our results. KT is supported by a Starting Grant from Trond Mohn Foundation (BFS2018REK01) and the University of Bergen. CAS receives financial support from European Union's Horizon 2020 research and innovation program under the grant agreement No. 82409; from Xunta de Galicia (Centro 
de investigacion de Galicia accreditation 2019-2022); from the European Union ERDF; from the Spanish Research State Agency by “Maria de Maeztu” Units of Excellence 
program MDM-2016-0692 and project FPA2017-83814-P and from the European Research Council project ERC-2018-ADG-835105 YoctoLHC. 
SPA and MS are supported by Grant Agency of the Czech Republic under grant 18-12859Y, by the Ministry of Education, Youth and Sports of the Czech Republic under grant 
LTT~17018, and by Charles University grant UNCE/SCI/013. SPA would also like to acknowledge Polish National Science Centre with grant no. DEC-2017/27/B/ST2/01985.
\end{acknowledgements}

\appendix
\section{Kinematical factors in splitting rates}
\label{sec:app:splitting_details}

The kinematical factors $\kappa_{ij}(z)$, appearing in the splitting rates, read
\begin{eqnarray}\label{eq:smallkappa-def}
\kappa_\glug(z)  &=& \sqrt{\frac{(1-z)C_A+z^2C_A}{z(1-z)}}\;,\\
\kappa_\qg(z) &=& \sqrt{\frac{C_F-z(1-z)C_{A}}{z(1-z)}}\;, \\
\kappa_\gq(z) &=& \sqrt{\frac{(1-z)C_A+z^2C_F}{z(1-z)}}\;, \\
\kappa_\qq(z)  &=& \sqrt{\frac{zC_A+(1-z)^2C_F}{z(1-z)}}\;. 
\end{eqnarray}
Finally,  the un-regularized Altarelli-Parisi splitting functions at leading order are given by
\begin{eqnarray}\label{eq:split-def}
P_\glug(z)  &=& \frac{1}{2} 2C_{A} \frac{[1-z(1-z)]^2}{z(1-z)}\;,\\
P_\qg(z) &=& \frac{1}{2} 2 n_f T_{R} \Big(z^2+(1-z)^2\Big)\;, \\
P_\gq(z) &=& \frac{1}{2} C_{F} \frac{1+(1-z)^2}{z}\;, \\
P_\qq(z)  &=& \frac{1}{2} C_{F}  \frac{1+z^2}{(1-z)}\;. 
\end{eqnarray}

\section{Multi-parton quenching in expanding medium}
\label{sec:fulljet-quenching}

Here we provide details about the calculation of the resummed jet quenching factor in an expanding medium. First, we have to calculate the phase space for vacuum-like emissions (VLEs) inside the quark-gluon plasma. The relevant time scale for decoherence can be read off the decoherence parameter, or the correlator of a dipole at a fixed opening angle,\footnote{In previous papers, the decoherence parameter is sometimes called $\Delta_\text{med}$, where $1-\Delta_\text{med} = S_{\rm dip}$ above.}
\beq
\label{eq:dip-factor}
S_{\rm dip}(t,0) = \rme^{- \int_0^t \rmd s\, v(\r(s),s)} \,,
\eeq
which corresponds to the probability for the dipole to remain in a color-correlated state. For our discussion, the dipole cross section $v(\r)$ is given simply by the harmonic oscillator approximation, i.e.
\beq
v(\r,s) = \frac{1}{4} \hat q(s) \r^2 \,,
\eeq
and the dipole follows the trajectory $\r(s) \simeq \theta_0 s$, where $\theta_0$ is the opening angle. We can define a decoherence time by considering the argument of the exponential in \eqref{eq:dip-factor}, namely
\beq
\frac{1}{4}  \theta_0^2 \int_0^t \rmd s\, \hat q (s) s^2 = \left(\frac{t}{\tdecoh} \right)^\alpha \,,
\eeq
where the index $\alpha$ depends on the density profile of the medium and follows from the resulting length dependence. As we will see below, $\alpha = 3$ for static and exponential profiles, while it is $\alpha = 3-\gamma$ for media whose density decays as $t^{-\gamma}$. 

Processes at time scales shorter than $\tdecoh$ are not resolved by the medium. Therefore the phase space for VLEs is given by the set of conditions \cite{Mehtar-Tani:2017web,Caucal:2018dla}
\beq
t_{\rm f} \ll t_{\rm d} \ll L 
\eeq
at DLA accuracy. At this accuracy, the available phase space for VLEs can be written as
\beq
\label{eq:ps-collimator}
\Pi_{\rm in} = 2 \frac{\alpha_s C_i}{\pi} \int_0^p \, \frac{\rmd \omega}{\omega} \int_0^R \frac{\rmd \theta}{\theta} \, \Theta({t_{\rm f } < t_{\rm d} < L}) \,,
\eeq
where the formation time is $\tform = 2/(\omega \theta^2)$, $C_i$ is the color charge of the parent parton and we treat the running coupling as a constant in this approximation. The quenching of a full jet is described by the collimator function \cite{Mehtar-Tani:2017web,Mehtar-Tani:2021fud}.
In the approximation of small energy losses, we use the linearized approximation to obtain
\beq
\label{eq:collimator-lin}
{\cal Q}_i(p_T,R) =  {\cal Q}^{(0)}_i(p_T) \exp\left[ \Pi_{\rm in} \left({\cal Q}_g^{(0)}(p_T) -1 \right) \right] \,,
\eeq
where $i =q, g$. We now proceed to work out in detail the scales of the phase space that allows for VLE's inside the medium for different medium profiles.

\paragraph{Static profile \& exponential profile :}

For these cases there is only a difference in a numerical constant.

The decoherence time is
\beq
\tdecoh = \left( \frac{4 {\cal N}}{\hat q_0 \theta^2} \right)^{1/3} \,,
\eeq
where
${\cal N} = 3$ for a static profile and ${\cal N} = e/(2e-5)$ for the exponential profile. The critical angle is then
\beq
\theta_c = \left(\frac{4{\cal N}}{\hat q_0 L^3} \right)^{1/2} \,.
\eeq
The phase space at DLA is
\beq
\Pi_{\rm in} = 2\bar \alpha \int_{\max[\theta_c,\theta_d, Q_0/\pT]}^R \frac{\rmd \theta}{\theta} \int_{\max [ (\theta_d/\theta)^{4/3},Q_0/(p\theta)]}^1 \frac{\rmd z}{z} \,,
\eeq
where
\beq
\theta_d= \left(\frac{2}{{\cal N}}\frac{\hat q_0}{p^3} \right)^{1/4} \,.
\eeq

\paragraph{Power-law profile :}
For a generic medium profile $\hat q(s) = \hat q_0(t_0/s)^\gamma$, we therefore find
\begin{align}
-\ln S_{q\bar q}(t,0) &= \frac{\theta_0^2 \hat q(L) L^3}{4} \frac{1 - \left(\frac{t_0}{L}\right)^{3-\gamma}}{3-\gamma} \,,\\
& = \left( \frac{L}{t_{\rm d}}\right)^{3-\gamma} \,,
\end{align}
where
\beq
t_{\rm d} = \left(\frac{4 (3-\gamma)}{\theta_0^2 \hat q(L) L^\gamma} \right)^{1/(3-\gamma)} \,,
\eeq
is the decoherence time \cite{Caucal:2020uic}. From the condition $t_{\rm d} = L$, we also deduce the characteristic decoherence angle
\beq
\theta_c = \left(\frac{4(3-\gamma)}{\hat q(L) L^3} \right)^{1/2} \,.
\eeq
The well known results for static medium, $\gamma = 0$, are straightforwardly reproduced. Furthermore, $\gamma = 1$ corresponds to the Bjorken expansion scenario.

The set of conditions for VLEs \cite{Mehtar-Tani:2017web,Caucal:2018dla}, $t_{\rm f} \ll t_{\rm d} \ll L$, demands that
\begin{align}
\omega^{3-\gamma} \theta^{4-2\gamma} &> \frac{2^{1-\gamma}}{3-\gamma} \hat q(L) L^\gamma \,,\\
\theta &> \theta_c \,.
\end{align}
In the Bjorken case, the first condition leads to a restriction on the transverse momentum of splittings, namely
\beq
k_\perp^2 \simeq \omega \theta > \frac{1}{2} \hat q_0 t_0 \,,
\eeq
combined with a restriction on the angle, given by $\theta > \theta_c = \sqrt{8/(\hat q_0 t_0 L^2)}$. Emissions that fail to meet these conditions are resolved in the medium, and should be treated as medium-induced processes. Finally, we also have to demand that the splittings are perturbative, i.e. $k_\perp > Q_0 \approx 1 $ GeV. 

The available phase space for VLEs at DLA accuracy can be written as
\beq
\label{eq:ps-collimator}
\Pi_{\rm in} = 2 \frac{\alpha_s C_i}{\pi} \int_{t_{\rm f } < t_{\rm d} < L} \, \frac{\rmd \omega}{\omega} \frac{\rmd \theta}{\theta} \Theta(R-\theta) \,,
\eeq
where $C_i$ is the color charge of the parent parton and we treat the running coupling as a constant in this approximation. Note, that for low-$p_T$ jets, i.e. when $p_T \leq \frac{1}{4}\hat q (L) L^2$, the requirement on the minimal angle becomes $\theta > \theta_d$, where
\beq
\theta_d = \frac{Q_s(L)}{p_T} \,,
\eeq
where $Q_s(L)^2 = \hat q(L) L/2$.
Changing variables in \eqref{eq:ps-collimator}, we finally obtain 
\begin{align}
\Pi_{\rm in} &= 2\frac{\alpha_s C_i}{\pi} \int_{R_{\rm min}}^R \frac{\rmd \theta}{\theta} \int_{k_{\perp,{\rm min}}}^{p_T \theta} \frac{\rmd k_\perp}{k_\perp} = \frac{\alpha_s C_i}{\pi} \,\ln\left( \frac{R}{R_{\rm min}}\right) \, \ln \left(\frac{p_T^2 R \,R_{\rm min}}{k^2_{\perp,{\rm min}}} \right) \,.
\end{align}
where $R_{\rm min} = \max[\theta_c,\theta_d]$ and $k_{\perp,{\rm min}} = \max[Q_s(L),Q_0]$. 
This phase space counts the number of VLE, created inside the medium, that will contribute to the energy loss of a jet.

\section{Nuclear modification factor with no nPDF effects}
\label{app:raa}

\begin{table}[t!]
\centering
\resizebox{\textwidth}{!}{ %
\begin{tabular}{|c||c|c||c|c||c|c||c|c||c|}
\hline
parameters          & \multicolumn{1}{c|}{$n_g$} & \multicolumn{1}{c||}{$n_q$} & \multicolumn{1}{c|}{$\beta_g$} & \multicolumn{1}{c||}{$\beta_q$} & \multicolumn{1}{c|}{$\gamma_g$} & \multicolumn{1}{c||}{$\gamma_q$} & \multicolumn{1}{c|}{$\delta_g$} & $\delta_q$   & $f_{q}(p_T = 40 GeV)$  \\ \hline
$ |y| < 0.3 $       & 3.21              & 2.52              & 0.99                 & 0.93                 & 0                             & 0           & 0                             & 0            & 0.16 \\ \hline 
$ |y| < 2.8 $       & 3.63              & 3.14              & 1.00                 & 0.92                 & 0                             & 0           & 0                             & 0        &  0.27    \\ \hline 
$ 2.1 < |y| < 2.8 $ & 5.63              & 5.00              & 0.35                 & 0.30                 & $-1.5 \cdot 10^{-10}$          & -0.19         & 0.24                  & 0.27  & 0.42 \\ \hline
\end{tabular} %
}
\caption{List of parameters from the extended power law characterization of input parton spectra for ``vacuum 1" configuration.}
\label{tab:inputs}
\end{table}
In this section we summarize analytic formulae for nuclear modification factor  for the case of not including effects from nuclear modifications of parton distribution 
functions. In that case we can directly propagate the extended power-law characterization of input power spectra (\ref{eq:MPL}) to quenching factor 
(\ref{eq:suppression-factor-2}) and flavor fraction (\ref{eq:fq}). The quenching factor reads then

\beq
\label{eq:app:suppression-factor-2}
{\cal Q}_i(\pT) = \int_0^1 \dd x \,(\pT)^{V_1} x^{n-1+V_2} D(x, \sqrt{x} \tau;\{i\}) \,,
\eeq
where,
\begin{eqnarray}
V_1&=&\beta_i \log(x)-\gamma_i \log^2(x)+2\gamma_i \log \Big(\frac{\pT}{\pTz}\Big)\log(x)\nonumber\\
&+&3\delta_i \log^2\Big(\frac{\pT}{\pTz}\Big)\log(x)-3\delta_i \log\Big(\frac{\pT}{\pTz}\Big)\log^2(x)+\delta_i \log^3(x)\\
V_2&=&\beta_i\log\Big(\frac{\pT}{x\pTz}\Big)+\gamma_i\log^2\Big(\frac{\pT}{x\pTz}\Big)+\delta_i\log^3\Big(\frac{\pT}{x\pTz}\Big).
\end{eqnarray} 
and flavor fraction is
\beq
\label{eq:app:fqpT}
f_{q}(\pT)= \left[ 1+\left(\frac{1-f_{q,0}}{f_{q,0}} \right) \left(\frac{\pTz}{\pT} \right)^{V_3} \right]^{-1} \,,
\eeq
where $f_{q,0}$ is fraction of quark-initiated jets at $\pTz$ and
\beq
V_3&=&{n_g-n_q}+(\beta_g-\beta_q)\log\bigg( \frac{\pT}{\pTz} \bigg)
+(\gamma_g-\gamma_q)\log^2\bigg( \frac{\pT}{\pTz} \bigg)\nonumber\\
&+&(\delta_g-\delta_q)\log^3\bigg( \frac{\pT}{\pTz} \bigg).
\eeq

These can be combined to $\Raa$ as follows

\beq
\label{eq:app:combRaa}
\Raa = f_q(\pT)\,{\cal Q}_q(\pT) + \big[1-f_q(\pT)\big] \, {\cal Q}_g(\pT) \,.
\eeq


\bibliographystyle{spphys}       
\bibliography{multiparton_expmed_v2}   
\end{document}